\documentclass[aps,nofootinbib]{revtex4}
\usepackage{color}
\newcommand{\sen}{\mbox{\rm sen}}
\usepackage{colordvi}

\usepackage{graphicx}

\begin{document}
\title{{\bf Explorando Sistemas Hamiltonianos II: Pontos de Equil\'{\i}brio Degenerados}\\
({\small Exploring Hamiltonian Systems II: Degenerate Equilibrium Points})}

\author{G. A. Monerat\footnote{E-mail: monerat@uerj.br},
E. V. Corr\^{e}a Silva, G. Oliveira-Neto, P. H. A. S. Nogueira, A. R.
P. de Assump\c{c}\~{a}o, T. M. G. de Oliveira}

\address{Departamento de Matem\'{a}tica e Computa\c{c}\~{a}o, 
 	Faculdade de Tecnologia, Universidade do Estado do Rio de Janeiro \\
	Rodovia Presidente Dutra, km 298, P\'{o}lo Industrial, 
	CEP 27537-000, Resende, RJ, Brasil.}

\begin{abstract}
Neste segundo artigo sobre sistemas Hamiltonianos, apresentamos o
m\'{e}todo da {explos\~{a}o ({\it blow-up})} para a determina\c{c}\~{a}o da natureza de pontos fixos (pontos de
equil\'{\i}brio) degenerados. Aplicamos o m\'{e}todo a dois modelos hamiltonianos com um e dois graus de liberdade, respectivamente. Primeiramente,
analisamos um sistema formado por um p\^{e}ndulo simples submetido a um torque externo constante $T$. Em seguida, consideramos um sistema
formado por um p\^{e}ndulo duplo com segmentos de comprimentos e massas iguais, tamb\'{e}m submetidos a torques externos constantes e n\~{a}o nulos. A presen\c{c}a de pontos de equil\'{\i}brio degenerados nos casos dos p\^{e}ndulos simples e duplo ocorre para certos valores dos torques externos.

\vspace{0.5cm}
In this second article on Hamiltonian systems, we present the {{\it blow-up}} method for the determination of the nature of 
{degenerate} fixed points (equilibrium points). {We apply the method to} two hamiltonian models with one and two degrees of 
{freedom, respectively.} Firstly we study a system formed by a simple pendulum submitted to a constant external torque $T$. 
Then we consider a system formed by a double pendulum of {segments with equal lengths and masses,} also 
{submitted to non-vanishing} constant external torques. The presence of degenerate equilibrium points in both cases of simple and double
pendulums {occurs for some values of the external torques.}
\end{abstract}

\maketitle

\newpage

\section{INTRODU\c{C}\~{A}O} 

\hspace{0.6cm}No estudo da din\^{a}mica de sistemas Hamiltonianos \cite{ozorio,ott}, a an\'{a}lise do espa\c{c}o de fase tem como ponto
de partida a busca de {\it pontos de equil\'{\i}brio} {(tamb\'em denominados {\it pontos fixos}).} A import\^{a}ncia de tais pontos 
{vem} do fato de que estes ``organizam'' a estrutura das \'{o}rbitas no espa\c{c}o de fase do {sistema (ou, em outras palavras, reduzem as possibilidades para esta estrutura).} {Na teoria dos sistemas din\^amicos \cite{alves}, um ponto de equil\'\i brio (ou, de modo equivalente, a estrutura dos espa\c co de fase nas suas vizinhan\c cas, que descreve o comportamento din\^amico do sistema nesta regi\~ao) pode ser classificado
mediante o processo de {\it lineariza\c{c}\~{a}o} \cite{ferrara}, que} consiste na expans\~{a}o {do sistema de equa\c c\~oes em s\'{e}rie de Taylor at\'{e} a primeira ordem, em} torno dos pontos de equil\'{\i}brio. {No entanto, quando} a matriz dos coeficientes cons\-tan\-tes do sistema linearizado em torno de um dado ponto de equil\'{\i}brio (matriz jacobiana) apresenta um ou mais autovalores nulos, {o processo de lineariza\c{c}\~{a}o n\~ao \'e suficiente para a caracteriza\c c\~ao do ponto de equil\'{\i}brio; este ent\~ao \'e dito
{\it degenerado} \cite{seydel}.}

{O matem\'{a}tico Ren\'{e} Thom \cite{rene} foi o primeiro a obter uma classifica\c{c}\~ao para pontos de equil\'{\i}brio de\-ge\-ne\-ra\-do.} 
 Uma classifica\c{c}\~{a}o muito simples {\'{e} tamb\'{e}m} apresentada por {\it Bautin} \cite{bau}. Em 1998, {\it Aranda} \cite{aranda} apresenta um
m\'{e}todo para an\'{a}lise de pontos de equil\'{\i}brio degenerados em  {sistemas planos (bi-dimensionais).} Um outro m\'{e}todo muito interessante \'{e} o da {explos\~{a}o (blow-up)} \cite{holmes}. Este m\'{e}todo tem sido aplicado na an\'{a}lise da din\^{a}mica de diversos sistemas f\'{\i}sicos 
{com pontos de equil\'{\i}brio degenerados.} Por exemplo, {\it Bogoyavlensky} \cite{bogo} utiliza o m\'{e}todo da explos\~{a}o em sistemas astrof\'{\i}sicos e {\it Monerat} \cite{teseMsc} em modelos cosmol\'{o}gicos de {Friedmann-Robertson-Walker,} com constante cosmol\'{o}gica e poeira,
para descrever a estrutura das curvas no espa\c{c}o de fase do modelo {pr\'{o}ximas \`a} singularidade. 

Recentemente {\it Monerat et al.} \cite{monerat7} apresentaram um estudo anal\'{\i}tico sobre o comportamento da din\^{a}mica 
{dos p\^{e}ndulos simples e duplo,} ambos submetidos a torques externos cons\-tan\-tes, na vizinhan\c{c}a dos pontos de equil\'{\i}brio 
{destes} sistemas. {Observa-se a presen\c{c}a de pontos de equil\'{\i}brio degenerados, para determinados valores dos torques externos.}
Para uma descri\c{c}\~{a}o completa desses sistemas, torna-se necess\'{a}rio a classifica\c{c}\~{a}o destes pontos de
equil\'{\i}brio degenerados, objetivo deste trabalho. Chamamos a aten\c{c}\~{a}o do leitor {para a restri\c c\~ao de nossa discuss\~ao a sistemas hamiltonianos; para o tratamento dos} demais sistemas, sugerimos a refer\^{e}ncia \cite{holmes}.

Neste trabalho, faremos uso do m\'{e}todo da explos\~{a}o para descrever a estrutura das curvas em uma vizinhan\c{c}a linear dos
pontos de equil\'{\i}brio degenerados existentes no espa\c{c}o de fase {dos p\^{e}ndulos} simples e duplo. {Na} se\c{c}\~{a}o \ref{metodo},
apresentamos o m\'{e}todo {da explos\~{a}o,} conforme exposto por {\it Guckenheimer} e {\it Holmes} \cite{holmes}, para um sistema hamiltoniano de um grau de liberdade, por quest\~{a}o de simplicidade. Na se\c{c}\~{a}o \ref{modelo}, {aplicamos} o m\'{e}todo para o 
{p\^{e}ndulo} simples submetido a um torque externo constante, {descrito} por uma fun\c{c}\~{a}o de Hamilton de um grau de liberdade. Na se\c{c}\~{a}o \ref{modelo2}, estendemos o m\'{e}todo para um sistema hamiltoniano de dois graus de liberdade formado por um p\^endulo duplo, tamb\'{e}m sujeito a
a\c{c}\~{a}o de torques externos constantes. Na se\c{c}\~{a}o \ref{conclusao}, {apresentamos} nossos coment\'{a}rios finais e conclus\~{o}es.

{\section{{O M\'ETODO DA EXPLOS\~{A}O (BLOW-UP)}}\label{metodo}}

{A id\'eia geral} {do m\'etodo da explos\~{a}o} \'e efetuar uma {transforma\c{c}\~ao} singular de coordenadas em torno do ponto de 
{equil\'\i brio} degenerado, a fim de {expand\'\i-lo} em uma esfera $n$-dimensional contendo
um n\'umero finito de pontos de equil\'{\i}brio do novo sistema. Por quest\~{a}o de simplicidade, apresentaremos nessa se\c{c}\~ao 
{o m\'etodo} da explos\~{a}o para sistemas hamiltonianos com um grau de {liberdade, estendendo-o em seguida para sistema}
sistemas com dois graus de liberdade. Uma descri\c{c}\~{a}o mais abrangente do m\'{e}todo pode ser vista {em} \cite{takens}. 

{Considere} um sistema descrito por uma {hamiltoniana} ${\cal H = H}(x,p_{x})$ cuja din\^{a}mica \'{e} governada pelas equa\c{c}\~{o}es
\begin{equation}
\frac{dx(t)}{dt}= \frac{\partial {\cal H}}{\partial p_{x}} = f(x,p_{x});\,\,\, \frac{dp_{x}(t)}{dt}= -\frac{\partial {\cal H}}{\partial x}= g(x,p_{x}),
\label{eq1}
\end{equation}

\noindent
em que $f(x,p_{x})$ e $g(x,p_{x})$ s\~{a}o fun\c{c}\~{o}es das coordenadas $\left\{x,\,\, p_{x} \right\}$. Suponha que exista no
espa\c{c}o  {de fase $\Phi$} do sistema pelo menos um ponto de equil\'{\i}brio ${\cal P}$ de coordenadas $x=x^{\star},\,\, p_{x}=p_{x}^{\star}$; ou
seja, um ponto tal que as derivadas de $x$ e $p_{x}$ em rela\c{c}\~{a}o ao par\^{a}metro temporal $t$ sejam nulas. Lembramos
que o ponto de equil\'{\i}brio em quest\~{a}o estar\'{a} situado sobre uma superf\'{\i}cie de energia no espa\c{c}o de fase, definida pela
equa\c{c}\~{a}o $E_{fixa}={\cal H}(x^{\star},\,p_{x}^{\star})$.

Se o ponto de equil\'{\i}brio $P(x^{\star},\,p_{x}^{\star})$ for degenerado {(ou seja, se a matriz jacobiana do sistema linearizado em torno de
$P(x^{\star},\,p_{x}^{\star})$ apresentar pelo menos um de seus autovalores nulos),} o comportamento na {vi\-zi\-nhan\-\c{c}a} linear desse
ponto n\~{a}o {poder\'a} ser determinado pelo processo de lineariza\c{c}\~{a}o do sistema, conforme discutido em \cite{monerat7}. 

{A estrutura das curvas em torno do ponto degenerado pode ser
analisada pelo m\'{e}todo da explos\~{a}o ({\it blow-up})
\cite{holmes}. Descrevemos os pontos $(x,y)$ 
uma vizinhan\c ca $\Phi$ do ponto de equil\'\i brio degenerado
$P(x^{\star},\,p_{x}^{\star})$ utilizando coordenadas polares $(r,
\theta)$, tal que $r \geq 0$, $0 \leq \theta < 2\pi$ e
\begin{eqnarray}
x-x^\star &=& r \cos\theta \\
p_{x} - p_x^\star &=&  r\ \sen \theta.
\label{eq2}
\end{eqnarray}
}
{\noindent Naturalmente, o caso $r=0$, para qualquer $\theta$,
corresponde ao ponto ${\cal P}$. Ap\'{o}s uma transforma\c c\~ao no
par\^{a}metro temporal, as equa\c c\~oes (\ref{eq1}) tomam a forma}
\begin{equation}
r\frac{\displaystyle dr(t)}{\displaystyle dt}= \bar{f}(r,\, \theta); \,\,\,\,\,
r\frac{\displaystyle d\theta(t)}{\displaystyle dt}= \bar{g}(r,\, \theta). \\
\label{eq3}
\end{equation}

{Podemos associar os pontos da vizinhan\c ca $\Phi$ a pontos de uma
superf\'\i cie cil\'\i ndrica $\Gamma$ (vide Fig. \ref{cilindro}), de
coordenadas (cil\'\i ndricas) $(r,\theta)$. Sejam $\Gamma_+$,
$\Gamma_0$ e $\Gamma_-$ os subconjuntos de $\Gamma$ tais que $r>0$,
$r=0$ e $r<0$, respectivamente; e seja $\Phi_+ = \Phi - \{{\cal P}\}$.
Cada ponto de $\Phi_+$ corresponde a um ponto de $\Gamma_+$, e
vice-versa. O ponto de equil\'\i brio degenerado ${\cal P}$
corresponde a $\Gamma_0$, ou seja, a circunfer\^encia $r=0$ sobre
$\Gamma$. N\~ao h\'a pontos em $\Phi$ correspondentes a pontos de
$\Gamma_-$.}

{Determinamos ent\~ao os pontos de equil\'{\i}brio do sistema (\ref{eq3}), 
${\cal Q}_i(r=0 \ ,{\theta}_i^{\star}) \in \Gamma_0$, tal que
$i=1,2,\dots,n$. Estes pontos correspondem \`as dire\c{c}\~{o}es, em
$\Phi$, segundo as quais as \'orbitas se aproximam ou se afastam do ponto degenerado
${\cal P}$.}
{Caso todos os pontos ${\cal Q}_i$ sejam n\~ao-degenerados, a
lineariza\c c\~ao do sistema (\ref{eq3}) em torno de cada um destes
determinar\'a completamente a estrutura das \'orbitas, nas suas
respectivas vizinhan\c{c}as lineares, e portanto a estrutura das
\'orbitas na vizinhan\c ca de ${\cal P}$. Contudo, se ainda houver
algum $Q_i$ degenerado, realiza-se uma nova explos\~ao em torno deste
ponto (introduzindo-se novas coordenadas), e assim por diante.
Sugerimos ao leitor interessado a Ref.\cite{holmes} para maiores
detalhes e exemplos do m\'etodo da explos\~ao.}

{\begin{figure}[h]
\includegraphics{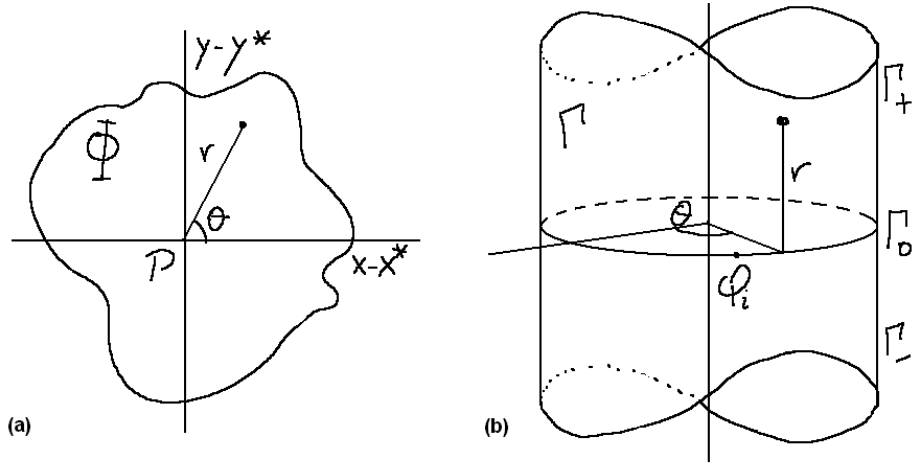}
\caption{
	{O m\'etodo da explos\~ao ({\it blow-up}). Em (a), pontos da vizinhan\c ca
	$\Phi$ de um ponto de equil\'\i brio degenerado $\cal P$ s\~ao
	descritos por coordenadas polares $(r,\theta)$ (vide Eq.(\ref{eq2}));
	as equa\c c\~oes de movimento s\~ao escritas nestas coordenadas (vide.
	Eq.(\ref{eq3})). Por sua vez, (b) mostra que $(r,\theta)$ podem ser interpretadas
	como coordenadas {\it cil\'\i ndricas} de pontos sobre uma superf\'\i
	cie cil\'\i ndrica $\Gamma$. Sejam $\Gamma_+$, $\Gamma_0$ e $\Gamma_-$
	os subconjuntos de $\Gamma$ tais que $r>0$, $r=0$ e $r<0$,
	respectivamente; e seja $\Phi_+ = \Phi - \{{\cal P}\}$. H\'a uma
	correspond\^encia biun\'\i voca entre os pontos de $\Phi_+$ e
	$\Gamma_+$; o ponto de equil\'\i brio degenerado ${\cal P}$
	corresponde a $\Gamma_0$ (a circunfer\^encia $r=0$ sobre $\Gamma$); e
	os pontos de $\Gamma_-$ n\~ao correspondem a nenhum ponto de $\Phi$.
	As equa\c c\~oes de movimento nas coordenadas $(r,\theta)$ ter\~ao
	pontos de equil\'\i brio ${\cal Q}_i \in \Gamma_0$. Se todos estes
	forem n\~ao-degenerados, a lineariza\c c\~ao em torno de cada ${\cal
	Q}_i$ determinar\'a a natureza das \'orbitas em torno de ${\cal P}$.
	Se houver ainda algum ${\cal Q}_i$ degenerado, uma nova explos\~ao em
	torno deste ponto ser\'a realizada, e assim por diante.	}}
\label{cilindro}
\end{figure}}

{\section{O P\^{e}ndulo Simples}\label{modelo}}

Vamos aplicar o m\'{e}todo descrito na se\c{c}\~{a}o \ref{metodo} 
	{ao p\^{e}ndulo simples} 
submetido a um torque externo constante de intensidade $T=mgL$, em que $m$ \'{e} a massa da
part\'{\i}cula, $L$ o comprimento do p\^{e}ndulo e $g$ 
{a acelera\c c\~ao da gravidade local.} 
{Tal sistema pode ser descrito por uma fun\c{c}\~{a}o de Hamilton na
forma}
\begin{equation}
H= \frac{p_{\varphi}^2}{2mL^2}-mgL\cos(\varphi)-mgL\varphi.
\label{eq4}
\end{equation}

As equa\c{c}\~{o}es de Hamilton que governam a din\^{a}mica desse
sistema apresentam um ponto de equil\'{\i}brio 
{no ponto $P^{*}(\varphi^*, p_\varphi^*)$, do espa\c co de fase, tal
que $\varphi^*= \frac{\pi}{2},\,\, p^*_{\varphi}=0$; este ponto possui}
{energia} $E_{fixa}=-\frac{mgL\pi}{2}$. 
{O processo de lineariza\c{c}\~{a}o \cite{monerat7}} das
equa\c{c}\~{o}es de Hamilton em torno 
{da} vizinhan\c{c}a linear do
ponto 
{$P^{*}$} 
coloca o sistema (\ref{eq3}) na forma
{\begin{equation}
\frac{d\vec{X}(t)}{dt}=J\left(\vec{X}(t)-\vec{X^{\star}}\right), 
\label{eq5}
\end{equation}}
\noindent em que 
{$\vec{X}(t)= (\varphi(t),\, p_{\varphi}(t))$, $\vec{X^{\star}}=
(\varphi^{\star},\, p_{\varphi}^{\star})$}
e $J$ \'{e} a 
{matriz jacobiana do sistema para o ponto $P^*$,}
{\begin{equation}
J=
\left(
\begin{array}{cc}
 0 & \frac{1}{\displaystyle mL^2}\\ 
 & \\
 0 & 0\\
\end{array}
\right). 
\label{eq6}
\end{equation}}
Os autovalores da matriz jacobiana $J$ s\~{a}o nulos, indicando 
{que o ponto de equil\'{\i}brio $P^*$ \'e um ponto de equil\'\i brio}
de\-ge\-ne\-ra\-do.

Vamos agora aplicar o m\'{e}todo apresentado na se\c{c}\~{a}o
\ref{metodo}. Para isso, 
{de acordo com (\ref{eq2}), faremos uma transforma\c c\~ao de vari\'{a}veis
\begin{equation}
\varphi = r \cos\theta; \,\,\,\,\, p_{\varphi}= r\ \sen \theta,
\end{equation}
tal que o}
sistema de equa\c{c}\~{o}es diferenciais
formado pelas equa\c{c}\~{o}es de Hamilton 
{assume} a forma

{\begin{eqnarray}
	r\frac{dr}{dt} &=& \frac{-r\cos(\theta)\left(m^2 gL^3\sen\left(r\sen(\theta)\right)-m^2 gL^3-r\sen(\theta)\right)}{mL^2},\label{eq7a}\\ \nonumber
	& & \\
	r\frac{d\theta}{dt} &=& \frac{m^2 gL^3\sen(r\sen(\theta))\sen(\theta)-m^2 gL^3\sen(\theta)-\sen(\theta)^2 r+r}{mL^2}.\label{eq7b}\\
	\nonumber
	\end{eqnarray}}

De acordo com o m\'{e}todo, inicialmente consideramos o limite 
{$r \rightarrow 0$ nas} equa\c{c}\~{o}es (\ref{eq7a}) e 
{(\ref{eq7b}); em seguida, determinamos}
os valores de $\theta=\theta^{\star}$ que fornecer\~{a}o as
coordenadas do ponto de equil\'{\i}brio degenerado nas novas
coordenadas. O resultado neste caso \'{e} $\theta^{\star}=0$. Tal
ponto est\'{a} associado a uma superf\'{\i}cie de energia
$\bar{E}_{fixa}=-mgL$. Ent\~{a}o, de acordo com o m\'{e}todo exposto
na se\c{c}\~{a}o \ref{metodo}, linearizamos o sistema em torno do
ponto 
{$(r=0,\, \theta^{\star}=0)$;} a matriz jacobiana nas novas
vari\'{a}veis \'{e}
\begin{equation}
\bar{J}=\left(
\begin{array}{ccc}
0 & mgL \\
& \\
-mgL & 1/mL^2\\
\end{array}
\right).
\label{eq8}
\end{equation}

\noindent A natureza de tal ponto de equil\'{\i}brio 
{\'e determinada pelos autovalores}  da matriz $\bar{J}$:
\begin{equation}
\lambda_{1} = \frac{1+\sqrt{1-4m^4L^6g^2}}{2mL^2}; \,\,\,\, \lambda_{2} = \frac{1-\sqrt{1-4m^4L^6g^2}}{2mL^2}.
\label{eq9}
\end{equation}

{Podemos observar que} se $m^4=\frac{1}{4L^6g^2}$ 
{ambos} os autovalores s\~{a}o reais e $\lambda_{1}=\lambda_{2}>0$,
indicando que tal ponto apresenta um equil\'{\i}brio 
{inst\'{a}vel, denominado  ``n\'o flexionado'' ({\it inflected node}) \cite{ferrara}.} 
A condi\c{c}\~{a}o $m^4>\frac{1}{4L^6g^2}$ indica que o ponto de equil\'{\i}brio \'{e} do tipo Foco, ou seja,	
{autovalores complexos da forma $\lambda = a\pm ib$,} em que $\left\{a,\, b\right\}$ s\~{a}o n\'{u}meros reais e positivos \cite{ferrara}. Esse
resultado indica que o volume do espa\c{c}o de fase n\~{a}o \'{e} conservado. Em sistemas Hamiltonianos conservativos, com dois graus de liberdade, pode ocorrer o aparecimento de pontos de equil\'{\i}brio do tipo Foco-Generalizado conforme mostrado em \cite{amaral}.
Nesses casos estes pontos 
{representam situa\c c\~oes de}
equil\'{\i}brio inst\'{a}vel. Para $m^4<\frac{1}{4L^6g^2}$ os
autovalores s\~{a}o reais e 
{distintos, configurando}  
um n\'{o} inst\'{a}vel.

A solu\c{c}\~{a}o geral do sistema linearizado formado pelas
equa\c{c}\~{o}es (\ref{eq7a}) e (\ref{eq7b}) em torno do ponto de
equil\'{\i}brio \'{e} uma superposi\c{c}\~{a}o (combina\c c\~ao
linear) das solu\c{c}\~{o}es linearmente independentes: 

{\begin{equation}
X_{i}(t) = \sum_{m=1}^{2} c^{(i)}_{m} A^{(i)}_{m}\ e^{\lambda^{(i)}_{m} t},
\label{eq10}
\end{equation}
}
\noindent
em que os $A^{(i)}_{m}$ s\~{a}o os autovetores associados aos
autovalores ${\lambda}^{(i)}_{m}$ da matriz $J_{i}$, e os coeficientes
$c^{(i)}_m$ s\~{a}o constantes de integra\c{c}\~{a}o que dependem das
condi\c{c}\~{o}es iniciais escolhidas.

Em termos das vari\'{a}veis $r,\, \theta$ as 
{solu\c{c}\~{o}es\cite{monerat7}}
v\'{a}lidas numa vizinhan\c{c}a linear do ponto de 
{equil\'{\i}brio  s\~{a}o da forma}
{\begin{eqnarray}
r(t) &=& c_{1} \lambda_1
e^{\lambda_1 \displaystyle  t} 
+c_{2}e^{\lambda_2 \displaystyle t}
\label{eq11} \\
\theta(t) &=& c_{3} \lambda_2
e^{\lambda_1\displaystyle  t} 
+c_{4}e^{\lambda_2\displaystyle t}.
\label{eq12}
\end{eqnarray}
}
	{em que $\lambda_1$ e $\lambda_2$ s\~ao dados por (\ref{eq9}).}
Como j\'{a} mencionado, as solu\c{c}\~{o}es (\ref{eq11}) e
(\ref{eq12}) dependem da rela\c{c}\~{a}o entre os valores da massa e
do comprimento do p\^{e}ndulo. A estrutura das curvas na
vizinhan\c{c}a linear dos pontos de equil\'{\i}brios para os casos de
n\'{o} inst\'{a}vel (quando $m^4<\frac{1}{4L^6g^2}$), 
{n\'o flexionado }
(quando $m^4=\frac{1}{4L^6g^2}$) e foco (quando
$m^4>\frac{1}{4L^6g^2}$) podem ser vistos na 
{Ref.\cite{ferrara}.}

\vspace{0.5cm}

{\section{O P\^{e}ndulo Duplo}\label{modelo2}}

{Um}  p\^{e}ndulo duplo de massas 
{id\^enticas} $m$ e 
{segmentos de comprimentos id\^enticos} $L$, 
{sob} a a\c{c}\~{a}o de torques externos constantes de
intensidades 
{$\beta_{1}= 2mgL$ e $\beta_{2}=mgL$, respectivamente,}
apresenta um ponto de equil\'{\i}brio degenerado no seu espa\c{c}o de
{fase\cite{monerat7}.} Este sistema \'{e} descrito por uma fun\c{c}\~{a}o de Hamilton
de dois graus de liberdade na forma
\begin{equation}
{\cal H} =\frac{\frac{p_{1}^2}{2}+p_{2}^2-\cos(\varphi_{1}-\varphi_{2})p_{1}p_{2}}{mL^2(1+\sen(\varphi_{1}-\varphi_{2})^2)}-2mgL\cos(\varphi_{1})-mgL\cos(\varphi_{2})+2mgL\varphi_{1}+mgL\varphi_{2}.
\label{eq13}
\end{equation}

As equa\c{c}\~{o}es de Hamilton governam a din\^{a}mica do 
{sistema; elas constituem}\
um conjunto de quatro equa\c{c}\~{o}es diferenciais n\~{a}o
lineares, 
{e possuem um} ponto de
equil\'{\i}brio degenerado 
{$P_0$  de} coordenadas
\begin{equation}
{P_0:}\,\,\, (p_{1}=0,\,\, \varphi_{1} = -\pi/2,\,\, p_{2}=0,\,\, \varphi_{2} = -\pi/2),
\label{eq14}
\end{equation}

\noindent
com uma energia associada 
{$E_0=-\frac{3}{2}\pi mgL$.}
Em se tratando de um ponto de equil\'{\i}brio degenerado, o processo
de lineariza\c{c}\~{a}o do sistema de equa\c{c}\~{o}es em torno desse
ponto 
{n\~ao \'e suficiente para caracteriz\'a-lo 
(i.e., caracterizar o comportamento do sistema em sua vizinhan\c ca).}
Para determinarmos a natureza deste ponto de
equil\'{\i}brio faremos uso do m\'{e}todo da 
{explos\~{a}o,} apresentado
na se\c{c}\~{a}o \ref{metodo}. 
{O p\^{e}ndulo duplo possui} dois graus de 
{liberdade $(\varphi_1,\varphi_2)$, aos quais associamos os momenta
canonicamente conjugados $(p_1,p_2)$, respectivamente. Portanto, utilizaremos} 
{coordenadas hiperesf\'ericas quadridimensionais, $(R,\, \theta,\,
\phi,\, \eta)$, definidas pelas transforma\c c\~oes \footnote{Uma
hiperesfera quadridimensional de raio $R$ \'e o lugar geom\'etrico dos
pontos do espa\c co euclidiano $\Re^4$, da forma
$(\varphi_1,p_1,\varphi_2,p_2)$, que guardam a mesma dist\^ancia $R$
da origem, satisfazendo portanto \`a equa\c c\~ao cartesiana
$\varphi_{1}^2+p_{1}^2+\varphi_{2}^2+p_{2}^2 = R^2$. Esta equa\c c\~ao \'e a generaliza\c
c\~ao do caso bidimensional (circunfer\^encia, $x^2+y^2=R^2$) e tridimensional (superf\'\i cie esf\'erica,
$x^2+y^2+z^2=R^2$).} de vari\'{a}veis}
{\begin{equation}
\left\{
\begin{array}{lll}
\varphi_{1}= R\ \sen(\theta)\ \cos(\phi)\ \sen(\eta) &;& p_{1}=R\ \sen(\theta)\ \sen(\phi)\ \sen(\eta),\\
& \\
\varphi_{2}= R\ \cos(\theta)\ \sen(\eta) &;& p_{2}=R\ \cos(\eta),\\
\end{array}
\right.
\label{eq15}
\end{equation}
}

{\noindent em que $\eta \in [0,\pi], \theta \in [0, 2\pi], \phi \in [0, 2\pi]$ 
e $R \in [0,\infty)$. Substituindo} 
as express\~{o}es (\ref{eq15}) nas equa\c{c}\~{o}es de
{Hamilton} do sistema obtido a partir da hamiltoniana
(\ref{eq13}), 
{obtemos} um novo conjunto de quatro equa\c{c}\~{o}es
diferenciais 
{nas} novas 
{vari\'{a}veis $(R, \theta, \phi, \eta)$. As equa\c{c}\~{o}es de Hamilton 
nas vari\'{a}veis originais pode ser visto na Ref. \cite{monerat7}; nas novas
vari\'aveis, as equa\c c\~oes tomam a forma}
\begin{equation}
\left\{
\begin{array}{lll}
R\frac{\displaystyle dR}{\displaystyle dt} &=& f_{1}(R,\theta,\phi,\eta);\,\,\,\,
R\frac{\displaystyle d\theta}{\displaystyle dt} = f_{2}(R,\theta,\phi,\eta)\\
& & \\
R\frac{\displaystyle d\phi}{\displaystyle dt} &=& f_{3}(R,\theta,\phi,\eta); \,\,\,\,
R\frac{\displaystyle d\eta}{\displaystyle dt} = f_{4}(R,\theta,\phi,\eta).\\
\end{array}
\right.
\label{eq16}
\end{equation}

\noindent
As coordenadas do ponto de equil\'{\i}brio degenerado nas novas
vari\'{a}veis s\~{a}o obtidas anulando o lado esquerdo das
equa\c{c}\~{o}es (\ref{eq16}), e em seguida, considerando o limite 
{$R\rightarrow 0$ das} fun\c{c}\~{o}es 
{$f_i$, $i \in \{1,2,3,4\}$.} Temos dessa forma um
sistema alg\'{e}brico de tr\^{e}s equa\c{c}\~{o}es (j\'{a} que a
primeira equa\c{c}\~{a}o \'{e} identicamente nula), 

{\begin{equation}
\left\{
\begin{array}{lll}
-2mgL\  {\cos(\phi)} \left[\sen(\theta)\sen(\eta)\right]^{-1}=0;\\
\\
-2mgL\ {\sen(\phi)\cos(\theta)} \left[\sen(\eta)\right]^{-1}=0;\\
\\ 
-2mgL\ \cos(\eta)\ \sen(\theta)\ \sen(\phi)+mLg\ \sen(\eta)=0,\\
\end{array}
\right.
\label{eq17}
\end{equation}}

\noindent
{cuja solu\c{c}\~{a}o fornece quatro pontos de equil\'{\i}brio,}
{\begin{equation}
\left\{
\begin{array}{lllllll}
Q_{1}:&\, \left(R=0,\, \phi =\pi/2,\, \eta = \arctan(2),\, \theta = \pi/2\right);\\
 & \\
Q_{2}:&\, \left(R=0,\, \phi = \pi/2,\, \eta = -\arctan(2),\,\theta = -\pi/2\right);\\
 & \\
Q_{3}:&\, \left(R=0,\,\phi = -\pi/2, \eta = -\arctan(2), \theta = \pi/2\right);\\ 
 & \\
Q_{4}:&\, \left(R=0,\, \phi = -\pi/2, \eta = \arctan(2), \theta = -\pi/2\right).\\
\end{array}
\right.
\label{eq18}
\end{equation}
}

Todos os pontos de equil\'{\i}brio descritos em (\ref{eq18}) est\~{a}o
associados a mesma superf\'{\i}cie de energia 
{\mbox{$\epsilon_{i}=-3mgL$.}}
{Ao linearizarmos} o sistema de equa\c{c}\~{o}es (\ref{eq16}) em torno de
{qualquer um dos pontos} de equil\'{\i}brio, 
{obtemos matrizes jacobianas id\^enticas, da forma}
\begin{equation}
J_{i}=\left(
\begin{array}{cccc}
-\sqrt{5}mgL & 0 & 0 & 0\\
0 & \sqrt{5}mgL & 0 & 0\\
-\frac{1}{2mL^2}& 0 & \sqrt{5}mgL & 0\\
0& 0& 0& \sqrt{5}mgL\\
\end{array}
\right).
\label{eq19}
\end{equation}

\noindent	
em que $i\in\{1,2,3,4\}$. O conjunto $\lambda_i$ de autovalores das matrizes 
{jacobianas}
fornecer\~{a}o a natureza  
{do i-\'esimo ponto. Obtemos,}
\begin{equation}
\lambda_{1,2}=\pm \sqrt{5}mgL;\,\,\, \lambda_{3,4}= \sqrt{5}mgL.
\label{eq20}
\end{equation}

{Este resultado indica \cite{ferrara} que} os quatro pontos de equil\'{\i}brio 
correspondem \`a combina\c{c}\~{a}o de uma sela ($\lambda_{1,2}$) e de um 
{n\'o flexionado} ($\lambda_{3,4}$); 
{representam, portanto, situa\c c\~oes de}
equil\'{\i}brio inst\'{a}vel.

\vspace{0.3cm}

{\section{CONCLUS\~{A}O E COMENT\'{A}RIOS FINAIS}\label{conclusao}}

Neste trabalho, aplicamos o m\'{e}todo da 
{explos\~{a}o ({\it blow-up})}
para a determina\c{c}\~{a}o da natureza de pontos de equil\'{\i}brio
degenerados no espa\c{c}o de fase de dois modelos 
{hamiltonianos: o}
p\^{e}ndulo simples 
{e o} p\^{e}ndulo 
{duplo, ambos sob a a\c c\~ao de torques externos constantes.}
{Conclu\'\i mos que {\it todos}}
os pontos degenerados existentes nestes dois sistemas descrevem 
{situa\c c\~oes de} equil\'{\i}brio inst\'{a}vel. No caso do
p\^{e}ndulo simples, mostramos que h\'{a} apenas um ponto de
equil\'{\i}brio 
{degenerado; este pode ser um n\'o inst\'avel simples ou flexionado,} 
dependendo da rela\c{c}\~{a}o entre a massa e o comprimento do p\^{e}ndulo. No caso
do p\^{e}ndulo duplo, 
{ap\'{o}s} a 
{explos\~{a}o} observamos a
exist\^{e}ncia de quatro pontos de 
{equil\'{\i}brio inst\'aveis} de mesma 
{natureza; cada um deles consistem no} 
produto direto de uma sela
hiperb\'{o}lica 
{(associada a um par de autovalores reais e sim\'etricos)}  por um
{n\'o flexionado.} Assim, o m\'{e}todo em quest\~{a}o
permite a descri\c{c}\~{a}o das solu\c{c}\~{o}es na vizinhan\c{c}a
linear de cada um dos pontos de equil\'{\i}brio degenerados, mostrando
ser uma t\'{e}cnica eficaz na an\'{a}lise da
estabilidade/instabilidade de tais sistemas. 
{A relativa}
simplicidade do m\'{e}todo 
{permite que este possa integrar o} programa de uma
disciplina de 
{Sistemas Din\^{a}micos} a ser 
{cursada} ap\'{o}s a
disciplina de 
{Mec\^{a}nica Anal\'{\i}tica\cite{nivaldo}}
nos cursos de gradua\c{c}\~{a}o em F\'{\i}sica, conforme discutido em
\cite{monerat7}.

\vspace{0.2cm}

\section*{AGRADECIMENTOS}

{G. A. Monerat, E. V. Corr\^{e}a Silva (Bolsista do CNPq, Brasil),
e G. Oliveira-Neto agradecem ao CNPq (Edital Universal CNPq/2006 -
Proc. 476852/2006-4). G.A. Monerat agradece a FAPERJ (Proc. No. E-
26/170.762/2004). T. M. G. de Oliveira (Bolsista do CNPq, Brasil) agradece ao 
CNPq pela bolsa de Inicia\c{c}\~{a}o Cient\'{\i}fica concedida no Edital PIBIC/2007.}

\section{Refer\^{e}ncias}

\end{document}